\documentstyle[aps,prl,epsf,floats]{revtex}

\begin{document}
  \twocolumn[%
  \hsize\textwidth\columnwidth\hsize\csname@twocolumnfalse\endcsname
  \title{Tunneling gap of laterally separated quantum Hall systems}
  \author{Marcus Kollar$^1$ and Subir Sachdev$^{1,2}$}
  \address{$^1$Department of Physics,
    Yale University, P.O. Box 208120,  New Haven, CT 06520-8120\\
    $^2$Department of Physics, Harvard University, Cambridge, MA 02138}
  \date{\today} 
  \maketitle
  \begin{abstract}
  We use a method of matched asymptotics to determine the energy
  gap of two counter-propagating, strongly interacting,
  quantum Hall edge states. The
  microscopic edge state dispersion and Coulomb interactions are
  used to precisely constrain the short-distance behavior of an
  integrable field theory, which then determines the low energy
  spectrum. We discuss the relationship of our results to the tunneling measurements
  of Kang {\em et al.}, Nature {\bf 403}, 59 (2000).
  \end{abstract}
  \pacs{PACS numbers:}
  ]

  For many solid-state materials, modern density-functional theory
  allows accurate computation of electronic properties {\em ab
    initio}. This success has not been extended to correlated electron systems
    like transition
  metal oxides, including the cuprate superconductors, and
  low-dimensional semiconductor heterostructures.
  In such systems, the reduced kinetic energy
  bandwidth enhances the importance of the correlations,
  and in some cases this leads to new quantum phases
  of matter. Quantum field-theoretic methods can often lead to a good
  understanding of the correlated phases, but their quantitative
  predictive power is limited: different energy scales can be related
  to each other, but at least one has to be measured experimentally. In
  this paper we shall describe one strongly correlated system in which
  it is possible to make a precise quantitative connection between the
  microscopic Hamiltonian and the electronic spectrum.

\begin{figure}[t]
    \epsfxsize=2.8in \centerline{\epsffile{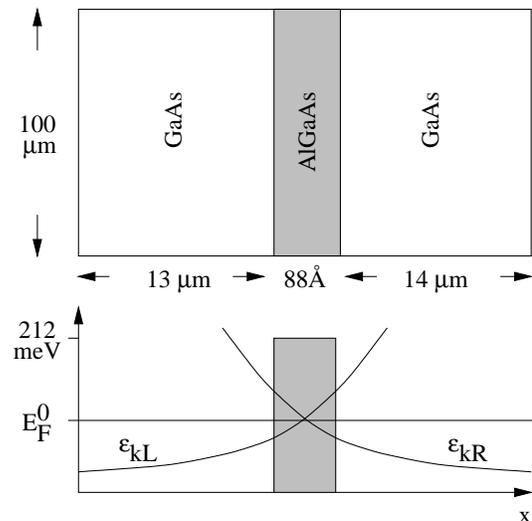}}
    \caption{The experiment of Kang {\em et
        al.}\protect\cite{kang}: two 2DEGs are laterally separated by
      an atomically precise tunneling barrier, with a height of
      212 meV\protect\cite{aditi}. The differential conductance is
      measured across the junction, with peaks occuring at zero bias
      for filling factors $\nu\approx1.2\text{-}1.5$ and
      $\nu\approx2.2\text{-}2.5$. The edge state dispersions are
      sketched for a barrier of infinite height; a gap appears for
      non-zero tunneling.\label{fig1}}
  \end{figure}

  We consider the recent experiments of Kang {\em et
    al.}\cite{kang} which studied tunneling between the
  counter-propagating edge states of two laterally separated quantum
  Hall states in two-dimensional electron gases (2DEG) in GaAs quantum
  wells (see Fig.~\ref{fig1}).
  The 2DEGs are separated by a smooth
  barrier so the elections in the edge states move
  ballistically along the barrier, while mixing via a
  weak matrix element for tunneling under the barrier, and strongly
  interacting via the Coulomb interaction. These coupled edge states
  constitute the correlated electron system of interest here.
  As has been argued by Kang {\em et al.}\cite{kang}, and more
  completely by Mitra and Girvin \cite{aditi}, the magnetic field and bias
  voltage dependence of the tunneling conductance act as sensitive
  probes of this system. In particular, the field
  interval over which a zero-bias peak is present determines
  the energy gap above the quantum ground state.

  This paper will make a microscopic computation of the energy gap
  of the coupled edge state system. We use the method
  of matched asymptotics: we combine the results of two
  separate calculations, valid on different length scales, over their
  intermediate regime of overlapping validity. It is the weak bare
  tunneling matrix element which leads to a clear-cut separation of
  scales crucial to our approach. At short scales, we will use a
  perturbative, microscopic calculation to account for the Coulomb
  interactions between the edge states (in the spirit of
  density-functional theory). At larger scales, we will use an
  integrable quantum field theory to determine the energy gap induced
  by the tunneling. However, the results of the quantum field theory
  depend sensitively on the nature of the cutoff regulating its {\em
    short} distance behavior: we will show how this is precisely
  specified by the {\em long} distance limit of the first calculation.
  The same quantum field theory has also been used in other recent
  theoretical works \cite{aditi,eric}; however, these works have used
  an ad-hoc specification of its cutoff. Our results show that there
  is no ambiguity in regulating the field theory, and that there is a
  precise matching procedure which connects it to the microscopic band
  structure.

  We consider the Hamiltonian $\hat{H} =
  \hat{H}_0 + \hat{H}_{\text{int}} + \hat{H}_t$.  For an infinitely
  high barrier the one-dimensional motion of the electrons on the two
  edge states ($\sigma=\text{L},\text{R}$) is represented by
  \cite{halperin,macdonald,ho}
  \begin{equation}
    \hat{H}_0
    =
    \sum_k \left(
      \epsilon_{k\text{L}}
      \hat{c}^{\dagger}_{k\text{L}}
      \hat{c}^{\phantom{\dagger}}_{k\text{L}} +
      \epsilon_{k\text{R}}
      \hat{c}^{\dagger}_{k\text{R}}
      \hat{c}^{\phantom{\dagger}}_{k\text{R}}
    \right)
    \,,\label{eq:H0}
  \end{equation}
  where $\hat{c}^{\dagger}_{k\sigma}$ is a creation operator for a
  fermion in an orbital with one-dimensional momentum, $k$, parallel
  to the barrier. The dispersion $\epsilon_{k\sigma}$ was determined
  by a solution of the Schr\"{o}dinger equation in the lowest Landau
  level, assuming complete spin polarization; we will need the full
  functional form of this dispersion, and not just a linearized Fermi
  velocity near the Fermi level. For zero bias voltage the dispersion
  satisfies
    $\epsilon_{k\text{L}}
    =
    \epsilon_{-k\text{R}}
    =
    \hbar\omega_{\text{c}}\,\nu_a(k)$,
  where $\omega_{\text{c}}$ is the cyclotron frequency. With $\ell$ the
  magnetic length, and $2a$ the width of the barrier,
  $\nu_a(k)$ depends on $k$ and $a$ only through
  $\nu_a(k)=w(k\ell+a/\ell)$.  The single-particle wavefunctions in
  the Landau gauge are $\psi^{\phantom{\ast}}_{k\sigma}(x,y) =
  \phi^{\phantom{\ast}}_{k\sigma}(x)\,e^{iky}/\sqrt{L}$, where
  $\phi_{k\sigma}(x)$ vanishes at the barrier; $L$ is the (effectively
  infinite) length of the barrier.

  The Coulomb interaction among the electrons in the edge states is
  \begin{equation}
    \hat{H}_{\text{int}}
    =
    \frac{1}{2L}
    \sum_{kk'q\sigma\sigma'}
    V^{\sigma \sigma'}_{k k' q}\,
    \hat{c}^{\dagger}_{k+q\sigma}
    \hat{c}^{\dagger}_{k'-q\sigma'}
    \hat{c}^{\phantom{\dagger}}_{k'\sigma'}
    \hat{c}^{\phantom{\dagger}}_{k \sigma}
    \,,
  \end{equation}
  where the matrix elements
  \begin{eqnarray}
    V_{kk' q}^{\sigma \sigma'}
    &=&
    \int\!dx_1\! \int\!dx_2\,
    \phi^{\ast}_{k+q\sigma}(x_1)
    \phi^{\phantom{\ast}}_{k\sigma}(x_1)
    V_{\text{C}}(q,x_1-x_2) \nonumber\\
    &~&~~~~~~~~~~~~~~~~~~~~~
    \phi^{\ast}_{k'-q\sigma'}(x_2)
    \phi^{\phantom{\ast}}_{k'\sigma'}(x_2)
    \label{eq:matrixelements}
  \end{eqnarray}
  are computed in terms of the wavefunctions, $\phi_{k\sigma}(x)$, of
  the edge states in the direction transverse to the barrier, with
  \begin{equation}
    V_{\text{C}}(q,x)
    =
    \frac{e^2}{\epsilon}
    \int\!dk\, \frac{e^{iqx}}{\sqrt{k^2+q^2}+q_{\text{TF}}}
    \,.\label{eq:kernel}
  \end{equation}
  Here $\epsilon$ is the dielectric constant of GaAs ($\epsilon=12.6$)
  and $q_{\text{TF}}$ is the Thomas-Fermi wavevector accounting for
  screening among the edge-state electrons. Finally, we write the
  tunneling between the edge states as
  \begin{eqnarray}
    \hat{H}_t
    &=&
    -t \int\!dx\, \hat{O}(x)\,,
    \\
    \hat{O}(x)
    &=&
    \frac{1}{L} \sum_{kk'}
    \hat{c}^{\dagger}_{k\text{L}}
    \hat{c}^{\phantom{\dagger}}_{k'\text{R}}
    \,e^{i (k-k')x}
    + \text{h.c.}
    \,.
  \end{eqnarray}
  In the absence of interactions this operator causes an energy gap of
  $\Delta_0=2t$ in the spectrum of $\hat{H}_0+\hat{H}_t$.

  As $t$ is very small compared to the other energy scales, it is
  useful to first consider the physics for $t=0$. The system
  $\hat{H}_0 + \hat{H}_{\text{int}}$ is a canonical example of a
  gapless Tomonaga-Luttinger (TL) liquid. One of its characterizations
  is the power-law correlator
  \begin{equation}
    \langle \hat{O}(x) \hat{O}(0) \rangle
    =
    \frac{2 A(\beta^2)^2 \Lambda^{2-4 \beta^2}}{|x|^{4 \beta^2}}
    \;\text{ for }\;
    x\rightarrow\infty,~t=0
    \,.\label{eq:TL-correlator}
  \end{equation}
  Here $\beta^2$ is a TL liquid exponent, and $A(\beta^2)$ is a
  dimensionless function; both quantities depend upon the complete
  microscopic details of $\hat{H}_0+\hat{H}_{\text{int}}$. The inverse
  length scale, $\Lambda$, can be chosen at our convenience: only the
  combination $A(\beta^2) \Lambda^{1-2 \beta^2}$ is determined by
  (\ref{eq:TL-correlator}), and any change in our choice for $\Lambda$
  will lead to a corresponding change in $A(\beta^2)$. None of our
  results will depend upon this choice. For the free particle case
  ($V_{\text{C}}=0$), a simple computation shows that $\beta^2=1/2$
  and $A(1/2) = 1/(2 \pi)$, independent of $\Lambda$.
  For $V_{\text{C}} \neq 0$, we computed $\beta$ and $A(\beta^2)$ both
  by perturbation theory in $V_{\text{C}}$ (valid for $\beta^2$ near
  $1/2$).  This was the time-consuming part of our calculation, and in
  principle, there is no problem in extending this computations to
  higher orders. For now, let us regard $\beta$ and $A(\beta^2)$ as
  known quantities (further details of their computation appear below)
  and proceed to a discussion of the consequences of $\hat{H}_t$.

  The effects of $\hat{H}_t$ are more easily discussed in a bosonized
  theory of the TL liquid \cite{stone}.  This is expressed in terms of
  the continuum, dimensionless, scalar field $\hat{\varphi} (x)$ with
  the Hamiltonian ($\hbar = 1$)
  \begin{equation}
    \hat{H}_{\text{TL}}
    =
    \frac{v_{\text{F}}}{2} \int\!dx \left[
      (\partial_x \hat{\varphi})^2 + \hat{\pi}^2
    \right],
  \end{equation}
  where $\hat{\pi}$ is the canonically conjugate momentum to
  $\hat{\varphi}$, and $v_{\text{F}}$ is the renormalized Fermi
  velocity. The standard bosonization mapping \cite{stone} also shows
  that the operator $\hat{O} (x)$ is proportional to
  $\cos(\sqrt{8\pi}\beta\hat{\varphi}(x))$, and so we conclude that
  the long-distance properties of the fermionic theory $\hat{H} =
  \hat{H}_0 + \hat{H}_{\text{int}} + \hat{H}_t$ are described by the
  sine-Gordon field theory with Hamiltonian
  \begin{equation}
    \hat{H}_{\text{sG}}
    =
    \hat{H}_{\text{TL}}
    -
    2 \alpha \int\!dx\,\cos(\sqrt{8\pi}\beta\hat{\varphi}(x))
    \,.\label{eq:SG-hamiltonian}
  \end{equation}
  This is a renormalizable and integrable quantum field theory, and
  all its properties are (in principle) determined in terms of $\beta$
  and $\alpha$ after suitable renormalization (normal-ordering) of the
  cosine interaction.  Following the convention of Zamolodchikov
  \cite{zamo}, we normalize the cosine interactions by
  \begin{eqnarray}
    \langle
    \cos (\sqrt{8 \pi} \beta \hat{\varphi}(x))
    \cos (\sqrt{8 \pi} \beta \hat{\varphi}(0))
    \rangle_{\text{sG}}
    =
    \frac{1}{2|x|^{4 \beta^2}}
    \,,\label{eq:SG-correlator}
  \end{eqnarray}
  for either $\alpha\to0$ and all $x$, or in the short-distance limit,
  $x\to0$ and all $\alpha$ \cite{lukyanov}.
  The theory (\ref{eq:SG-hamiltonian}) contains a soliton and an
  antisoliton with mass $M$, as well as bound states (``breathers'')
  with masses $m_n = 2M\sin(\pi n\xi/2)$, $n = 1,2,\ldots,<1/\xi$,
  where $\xi = \beta^2/(1-\beta^2)$, and $\alpha$ and $M$ are related
  by \cite{zamo}
  \begin{equation}
    \frac{\alpha}{v_{\text{F}}}
    =
    \frac{\Gamma(\beta^2)}{\pi \Gamma(1-\beta^2)}
    \left[
      \frac{M}{v_{\text{F}}}
      \frac{
        \sqrt{\pi}\,\Gamma({\textstyle\frac{1+\xi}{2}})}{
        2\,\Gamma({\textstyle{\frac{\xi}{2}}})}
    \right]^{2-2\beta^2}
    \,.\label{eq:SG-ratio}
  \end{equation}
  Before we can apply (\ref{eq:SG-ratio}) we need to determine $\alpha$.
  It is tempting to do this by using the standard representation of
  the fermionic fields in terms of exponentials of $\hat{\varphi}$ and
  $\hat{\pi}$ in the expression for $\hat{O}(x)$, and thus map
  $\hat{H}_t$ to the cosine term in $\hat{H}_{\text{sG}}$.  However,
  this is incorrect: this mapping implicitly assumes that the fermions
  are free ($V_{\text{C}}=0$), and does not properly account for the
  essential, non-universal renormalization of the overall scale of
  $\hat{O}(x)$. Instead, the proper answer is obtained with the
  realization that the results (\ref{eq:TL-correlator}) and
  (\ref{eq:SG-correlator}) in fact refer to the same regime of $x$:
  the long-distance limit of $\hat{H}_0+\hat{H}_{\text{int}}$ is
  described by the TL liquid behavior in (\ref{eq:TL-correlator}), and
  this co-incides with the short-distance limit of the continuum
  theory $\hat{H}_{\text{sG}}$ as described by the TL liquid behavior
  in (\ref{eq:SG-correlator}). In this manner we may deduce the
  proportionality constant between $\hat{O} (x)$ and $\cos (\sqrt{8
    \pi} \beta \hat{\varphi} (x))$, and so obtain
  \begin{equation}
    \alpha = t A(\beta^2) \Lambda^{1-2\beta^2}
    \,.\label{eq:prop-const}
  \end{equation}
  With the knowledge of $\beta$ and $A(\beta^2)$ the energy gap,
  $\Delta$, can be obtained from
  (\ref{eq:SG-ratio})--(\ref{eq:prop-const}), via
  \begin{equation}
    \Delta
    =
    2\,\text{min}(m_1,M)
    \,.\label{eq:SG-gap}
  \end{equation}
  In particular $\Delta=\Delta_0$ for the non-interacting case,
  $\beta^2=1/2$, as expected.

  Before we discuss our final numerical results for $\Delta$, let us
  present a few details of the promised computation of $\beta$ and
  $A(\beta^2)$. An expansion of the TL correlator
  (\ref{eq:TL-correlator}) for $\beta^2$ near $1/2$ yields
  \begin{eqnarray}
    \lefteqn{
      \langle\hat{O}(x)\hat{O}(0)\rangle
      =
      \frac{1}{2\pi^2x^2}
      \;\times }
    &&
    \nonumber\\
    &~&~~~~~
    \left[
      1+
      (\beta^2-{\textstyle\frac{1}{2}})\left(
        4\pi A'({\textstyle\frac{1}{2}})-4\ln(x\Lambda)
      \right)
      +\cdots
    \right].\label{eq:TL-correlator-expanded}
  \end{eqnarray}
  On the other hand we calculate the correlation function of the
  tunneling operator in the fermionic theory with $t=0$ to first order
  in $\hat{H}_{\text{int}}$,
  \begin{eqnarray}
    \langle\hat{O}(x)\hat{O}(0)\rangle
    &=&
    \frac{1}{2\pi^2x^2}
    \left[
      1+
      \frac{e^2/(\epsilon\ell)}{\hbar\omega_{\text{c}}}
      \,
      \frac{g(x/\ell)}{\pi}
      +\cdots
    \right]
    \,,\label{eq:PT-correlator}
  \end{eqnarray}
  and extract the asymptotic behavior of $g(x)$,
  \begin{eqnarray}
    g(x)
    &=&
    c_1\ln x+c_2+O(x^{-2})
    \;\text{ for }\;
    x\rightarrow\infty
    \,.\label{eq:PT-asymptotics}
  \end{eqnarray}
  Comparison with (\ref{eq:TL-correlator-expanded}) leads to
  expressions for $\beta^2$ and $A'(\frac{1}{2})$ in terms of
  $c_1$ and $c_2$.
  It remains to determine $c_1$ and $c_2$ from $g(x)$.
  It is straightforward to calculate the
  required correlation function in terms of a two-particle Green
  function
  \begin{eqnarray}
    \lefteqn{
      \langle\hat{O}(x)\hat{O}(0)\rangle
      =
      }
    &&
    \nonumber\\
    &&
    \sum_{kk'q\sigma\sigma'}
    e^{iqx}
    \langle
    \text{T}
    \hat{c}^{\phantom{\dagger}}_{k'\sigma'}     (\tau_2)
    \hat{c}^{\phantom{\dagger}}_{k \bar{\sigma}}(\tau_3)
    \hat{c}^{\dagger}_{k+q\sigma}               (\tau_1)
    \hat{c}^{\dagger}_{k'-q\sigma'}             (\tau_4)
    \rangle
    \,,\label{eq:PT-green}
  \end{eqnarray}
  for vanishing imaginary times, with $\tau_1$ $>$ $\tau_2$ $>$
  $\tau_3$ $>$ $\tau_4$. Up to first order in $V_{\text{C}}$ we obtain
  \begin{equation}
    \frac{g(x)}{x^2}
    =
    \int_{ 0}^{\infty}\!\!\!\!\!dq
    \int_{-q}^{0}     \!\!\!\!dk
    \int_{ 0}^{q}     \!\!\!\!dk'
    \,
    F(k,k',q)\cos(k+q-k')x
    ,\!\!\label{eq:PT-contribution}
  \end{equation}
  where $F$ and its arguments are dimensionless,
  \begin{equation}
    \frac{e^2/(\epsilon\ell)}{\hbar\omega_{\text{c}}}
    F(k\ell,k'\ell,q\ell)
    =
    \frac{V_{kk'q}^{\text{LR}}}{
      \epsilon_{k+q\text{L}}  - \epsilon_{k\text{L}}+
      \epsilon_{k'-q\text{R}} - \epsilon_{k\text{R}}}
    .\!\!\label{F-function}
  \end{equation}
  It is difficult to determine the logarithmic asymptotics numerically
  from this integral over an oscillatory function. We therefore
  consider the Mellin transform $G(s)$ of $g(x)$, which is defined by
  \begin{eqnarray}
    G(s)
    &=&
    \int_1^{\infty}\!dx
    \,x^{s-1}\,g(x)
    \,,\;\;\;s<0
    \,.\label{eq:Mellin-def}
  \end{eqnarray}
  The following particular transform is important for our purposes,
  \begin{eqnarray}
    \int_1^{\infty}\!dx
    \,x^{s-1}\,\frac{(\ln x)^n}{x^m}
    &=&
    \frac{(-1)^{n+1}n!}{(s-m)^{n+1}}
    \,,
  \end{eqnarray}
  which shows that powers of logarithms transform into poles in the
  Laurent expansion of $G(s)$. In the present case we expect that
  \begin{eqnarray}
    G(s)
    &=&
    \frac{c_1}{s^2}-\frac{c_2}{s}+O(s^0)
    \text{ for }s\to0
    \,.\label{eq:Mellin-asympt}
  \end{eqnarray}
  Performing two partial integration w.r.t $q$ in
  (\ref{eq:PT-contribution}) leads to the function $\tilde{g}(p)$, a
  Fourier-like representation of $g(x)$,
  \begin{eqnarray}
    g(x)
    &=&
    \int_{0}^{\infty}\!\!\!\!dp
    \,(1-\cos px)\frac{\partial^2\tilde{g}(p)}{\partial p^2}
    \,,
    \\
    \tilde{g}(p)
    &=&
    \int_{0}^{\infty}\!\!\!\!dk
    \int_{0}^{\infty}\!\!\!\!dk'
    \sum_{\lambda=\pm1}\!
    F(-\lambda k,\lambda k',\lambda (k+k'+p))
    ,\!\!
  \end{eqnarray}
  on which the integration in (\ref{eq:Mellin-def}) can be performed,
  with the result (primes mean derivatives w.r.t. $p$)
  \begin{eqnarray}
    G(s)
    &=&
    \int_{0}^{\infty}\!\!dp
    \left[-\frac{\Gamma(s)\cos\frac{\pi s}{2}}{p^s}
    \right.
    \nonumber\\
    &&~~~~~~~~~~~
    \left.
    +\frac{{}_1F_2(\frac{s}{2},1+\frac{s}{2},\frac{1}{2},-\frac{p^2}{4})-1}{s}
    \right]\tilde{g}''(p)
    \\
    &=&
    \frac{\gamma s-1}{s}
    \int_0^{\infty}\!dp
    \,p^{-s}
    \,\tilde{g}''(p)
    +O(s^0)
    \,,
  \end{eqnarray}
  where $\gamma\approx0.577$ is Euler's constant.  For $p\to0$ it can
  be shown that $\tilde{g}''(p)\propto
  V_{000}^{\text{LR}}/(v_{\text{F}}^0p)$; a partial integration then
  yields
  \begin{eqnarray}
    G(s)
    &=&
    \frac{\gamma s-1}{s^2}
    \int_0^{\infty}\!dp
    \,(1-s\ln p)
    \,(p\tilde{g}''(p))'
    +O(s^0)
    \,,\!\!
  \end{eqnarray}
  from which we obtain, by comparison with (\ref{eq:Mellin-asympt}),
  and some rearrangement,
  \begin{eqnarray}
    c_1
    &=&
    \lim\limits_{p\to0}\;
    p\,\tilde{g}''(p)
    =
    \frac{\omega_{\text{c}}\ell}{v_{\text{F}}^0}
    \frac{V_{000}^{\text{LR}}}{e^2/(\epsilon\ell)}
    ,\,\label{eq:c1}
    \\
    \frac{c_2}{c_1}
    &=&
    \int_0^{\infty}\!\!\!\!dp
    \left[
      \frac{\tilde{g}''(p)}{c_1}-\frac{e^{-p}}{p}
    \right]
    ,\,\label{eq:c2}
  \end{eqnarray}
  where an integral representation for $\gamma$ was used to subtract
  the pole at the origin.  The integral in (\ref{eq:c2}) can be
  performed, with the result
  \begin{eqnarray}
    \frac{c_2}{c_1}
    &=&
    \lim\limits_{p\to0}
    \int_{0}^{\infty}\!\!\!\!dk
    \int_{0}^{\infty}\!\!\!\!dk'
    \frac{\partial}{\partial p}
    \left[
      \frac{e^{-(k+k'+p)}}{k+k'+p}
    \right.
    \nonumber\\
    &&~~~~~~~
    \left.
      -
      \sum_{\lambda=\pm1}\!
      \frac{
        F(-\lambda k,\lambda k',\lambda (k+k'+p))
        }{c_1}
    \right]
    ,\,\label{eq:c2integrated}
  \end{eqnarray}
  which we evaluate numerically. To obtain the matrix elements
  (\ref{eq:matrixelements}) we calculated the integral in
  (\ref{eq:kernel}), which is more easily evaluated in terms of a
  Bessel function, ($a,b>0$)
  \begin{eqnarray}
    \int_{0}^{\infty}\!\!\!\!\frac{dp\,\cos{p}}{\sqrt{p^2+a^2}+b}
    =
    \int_{0}^{\infty}\!\!\!\!dx\,
    \frac{ax\,e^{-bx}}{\sqrt{1+x^2}}
    K_1({a\sqrt{1+x^2}})
    \,,\!\!\label{eq:f-integral}
  \end{eqnarray}
  and obtained the real-space wavefunctions of the lowest Landau level
  edge states from the solution of the Schr\"{o}dinger equation. This
  completes the calculation of $\beta$ and $A'(1/2)$.

  Finally, the knowledge of $c_1$ and $c_2$ allows determination of the
  energy gap. Eliminating $\alpha$
  from (\ref{eq:SG-ratio})--(\ref{eq:prop-const}), and expanding in
  $\beta^2-\frac{1}{2}$, we obtain
  \begin{equation}
    \frac{M}{t}
    =
    1+2(\beta^2-{\textstyle\frac{1}{2}})
    \left[
      \pi A'({\textstyle\frac{1}{2}})
      +\ln\frac{e^\gamma t}{2v_{\text{F}}\Lambda}
    \right]
    +\cdots
    \,.
  \end{equation}
  To this order, it is permissible to use the bare Fermi velocity
  $v_{\text{F}}^0=\omega_{\text{c}}\ell\,w'\!(a/\ell)$. Matching the expansions
  in (\ref{eq:TL-correlator-expanded}) and (\ref{eq:PT-asymptotics}) to express
  the results in terms of $c_{1,2}$, the final
  result for the energy gap in the presence of interactions,
  $\Delta=2M$, is
  \begin{eqnarray}
    \frac{\Delta}{\Delta_0}
    &=&
    1
    +
    \frac{c_1}{2\pi}\frac{e^2/(\epsilon\ell)}{\hbar\omega_{\text{c}}}
    \left[
      \ln\frac{4\hbar v_{\text{F}}^0}{\Delta_0\ell}
      +\frac{c_2}{c_1}-\gamma
    \right]
    \,,\label{eq:Delta-result}
  \end{eqnarray}
  where the microscopic scale $\Lambda$ has cancelled.  Note that only
  microscopic parameters appear in this equation; except for the
  tunneling $t$ itself (which enters via the bare gap, $\Delta_0=2t$),
  they all depend only on properties of the two 2DEGs without
  tunneling. It is also obvious that $t$ enters into $\Delta$ in a
  non-perturbative fashion.

  Our results for $\Delta$ are shown in Table~\ref{tab1}.
 \begin{table}
    \begin{tabular}{ccccccc}
      $2a/l$ &
      $2a$ {[\AA]} &
      $v_{\text{F}}^0/(\omega_{\text{c}}\ell)$ &
      $\Delta_0$ {[K]}
                      &
      $q_{\text{TF}}\ell$ &
      $2\beta^2$ &
      $\Delta$ {[K]}
      \\ \hline
      0.84 &
      88 &
      1.47 &
      0.52 &
      0.05 &
      0.64 &
      1.49
      \\
      0.84 &
      88 &
      1.47 &
      0.52 &
      0.10 &
      0.76 &
      1.19
      \\
      0.50 &
      52 &
      1.33 &
      4.00 &
      0.05 &
      0.58 &
      9.25
      \\
      0.50 &
      52 &
      1.33 &
      4.00 &
      0.10 &
      0.72 &
      7.83
    \end{tabular}
    \vspace{0.1in}
    \caption{Numerical results for the correlated tunneling gap $\Delta$
      [Eq.~(\protect\ref{eq:Delta-result})] and the TL liquid exponent
      $2\beta^2$, obtained for a magnetic
      field of $B=6\text{~T}$, corresponding to magnetic length
      $\ell=105\text{~\AA}$ and cyclotron energy
      $\omega_{\text{c}}=10.4\text{~meV}$
      (based on the effective mass $m^*=0.067m$ for GaAs), and several
      values of the barrier width and Thomas-Fermi screening vector
      $q_{\text{TF}}$. The barrier width determines the tunneling gap
      $\Delta_0$ in the absence of interactions and was calculated in
      Ref.~\protect\cite{aditi}.\label{tab1}}
  \end{table}
  We find that
  the correlated gap is larger by a factor of 2-3 than the bare gap.
  Compared with the Hartree-Fock results of Mitra and
  Girvin\cite{aditi} our results for $\Delta$ are larger by about
  10~\% for the experimental barrier width of $88\text{~\AA}$ and by
  about 50~\% for a smaller barrier of $52\text{~\AA}$. We conclude
  that the gap is significantly enhanced by Coulomb correlations
  especially for strong tunneling.

  The gap can be related to the filling factor range
  $\nu=\nu^*\pm\delta\nu$ where a zero-bias peak is observed in the
  differential conductance\cite{kang,aditi}; this signal is expected
  when the Fermi energy falls inside the gap.  From our results its
  appearance is predicted in a range of
  $\delta\nu\approx\Delta/(\hbar\omega_{\text{c}})\approx0.01$, much
  smaller than the experimentally observed $\delta\nu\approx0.15$. At
  present the reason for this difference is unclear. It may be due to
  disorder in the sample, which has been suggested to further increase
  the gap\cite{aditi}.  Another puzzle is that for the non-interacting
  system the two branches of the dispersion are predicted to cross
  above the second Landau level, so that the zero-bias conductance
  peak is expected near $\nu^*\approx2$, but is found experimentally
  at $\nu^*\approx1$ (our approach does not provide an estimate of
  $\nu^*$).  In any case it may be worthwhile to repeat the experiment
  with a narrower barrier, since in this case the larger bare
  tunneling amplitudes and Coulomb matrix elements should make
  correlations dominate over possible disorder effects.

  In conclusion, we have provided a microscopic and quantitative computation
  of an observable in a correlated electron system,
  the energy gap for lateral tunneling between two
  interacting quantum Hall systems, and compared our result to
  experiment. Our method of matched asymptotics should find
  applications in other correlated systems.

  We thank S.~Girvin, A.~Mitra, and W.~Kang for useful discussions.
  This research was supported by US NSF Grant DMR 00--98226. M.~K.\
  gratefully acknowledges support from the Deutsche
  Forschungsgemeinschaft.

\end{document}